\documentclass[useAMS,usenatbib]{mn2e}
\usepackage{amssymb}
\usepackage{longtable}
\usepackage{hyperref}
\usepackage{graphicx}
\usepackage{subfigure}
\usepackage{color}
\usepackage{ulem}
\usepackage{amsmath}
\usepackage{stmaryrd}

\title[Centre offsets v.s. velocity dispersion]
   {An anti-correlation between barycentre offsets and velocity dispersion for highest overdensity subhalos in cosmological simulations}
\author[Ming-Hua Li \& Hai-Nan Lin]
   {Ming-Hua~Li$^1$\thanks{E-mail: liminghua@mail.sysu.edu.cn}
  and Hai-Nan~Lin$^2$
  \newauthor \\
  $^1$School of Physics and Astronomy, Sun Yat-Sen University, 135 Xingang Xi
      Road, Guangzhou 510275, China\\
  $^2$Department of Physics, Chongqing University, Chongqing 401331, China
  }
\date{Draft version}

\pagerange{\pageref{firstpage}--\pageref{lastpage}} \pubyear{2017}

\def\LaTeX{L\kern-.36em\raise.3ex\hbox{a}\kern-.15em
    T\kern-.1667em\lower.7ex\hbox{E}\kern-.125emX}

\def\bc{\begin{centre}}
\def\ec{\end{centre}}
\def\be{\begin{eqnarray}}
\def\ee{\end{eqnarray}}

\begin{document}

\label{firstpage}

\maketitle

\begin{abstract} 
{In this paper, we use two hydrodynamic simulations to study the barycentre offsets $r_{{\rm off}}$ between the entire halo and the gas or dark matter subhalo with largest overdensity. We restrict our analysis to those halos with large rescaled offsets, i.e. $r_{{\rm off}}/r_{200}\gtrsim 0.05$. 
The halos which are less massive and at higher redshifts are more likely to have larger $r_{{\rm off}}/r_{200}$. The 3D velocity dispersion of the subhalos with largest overdensity is found to have a similar mass and redshift dependence as $r_{{\rm off}}/r_{200}$. We also find that the gas subhalos with maximum overdensity and larger gas velocity dispersion $\sigma_{v}^{\rm ICM,sub}/\sigma_{200}$ tend to have smaller barycentre offsets $r_{{\rm off}}/r_{200}$. Similar result is found for the highest overdensity dark matter subhalos but with less significance. This anti-correlation is more significant for the snapshot $z=0$ than $z=0.2$ and $0.5$. The underlying physical mechanism for this anti-correlation remains to be explored.
 }
\end{abstract}


\section{Introduction}
In the current structure formation theory, all matter structures in our Universe, like filaments, voids, and clusters of galaxies, etc., have their origins in the primordial density fluctuations during inflation in the early Universe. In the cosmic evolution, these small fluctuations in the matter density field are later amplified by gravity, giving rise to dark matter halos with different mass and sizes. The galaxies and clusters of galaxies in our Universe are believed to form in these dark halos at later times. This scenario predicts that ordinary matter (baryonic matter) should follow tightly the distribution and evolution of dark matter if one neglects the baryonic feedbacks.

On the other hand, the spatial offsets between the baryons and the dark matter in galaxies and clusters of galaxies have actually been well observed and studied in the past few years.
After studying the projected offsets between the dominant component of baryonic matter centre and the gravitational centre of 38 galaxy clusters, \citet{Shan2010} found that 45 per cent of them have offsets $> 10$ arcsec. \citet{AM2011} reported that the fraction of miscentred clusters in the maxBCG catalogue (see \citet{Koester2007}) is $\sim 28$ per cent, which agrees with what were found by \citet{john07a, john07b} and \citet{HW2010} from their cosmological simulations. Later works have also provided concrete evidences of these offsets
\citep{Sanderson2009, Hudson2010, oguri2010, Mann2012, george2012, zitrin2012, Rossetti2016}.

Although the evidences are apparent, the underlying mechanism for these offsets is still not fully understood. For galaxies and clusters of galaxies, the baryon physics such as heating, radiative cooling, star formation, supernovae and AGN feedbacks, could actually lead to the spatial offsets between the baryonic components (i.e. the BCGs and the ICM) and dark matter. Other kinematic process such as merging and collision (i.e. the Bullet Cluster 1E0657-558, see \citet{Bradac2006, Clowe2006}) would also contribute to the separation of baryons and dark matter in clusters of galaxies. Moverover, \citet{Liao2016} recently showed that the dark matter and gas content of a halo could be initially segregated even in the absence of radiative cooling. Knowledge of the offsets between baryonic and dark matter is not only important for an accurate interpretation of cluster properties but also essential for reducing systematic errors when using clusters of galaxies as cosmological probes \citep{mandelbaum2010, rozo2011, viola2015}.

It is interesting to investigate whether the spatial offsets between baryons and the dark matter in galaxies and clusters of galaxies could be related to some intrinsic properties of the system. In this paper, we perform two hydrodynamical simulations to examine explicitly this assumption. Mass and redshift dependence of the offsets are explored. 
If such a relation exists, it may provide an alternative way to estimate the offsets and helps to shed lights on the underlying physics that are responsible for the offsets.

The rest of the paper is organized as follows. Section 2 is about the simulations we used. In Section 3, we study the mass and redshift evolution of the rescaled barycentre offsets between the gas and dark matter subhalos and the entire system. The correlation between the gas velocity dispersion and the offsets are also explored in this section. Conclusions and discussion about the results are presented in Section 4.

\section{Numerical Simulations}
Our study is based on two hydrodynamic simulations started from $z_{\rm ini} = 120$, which are performed by the Tree-PM $N$-body/SPH code GADGET-2 \citep{springel2005}. The fiducial simulation contains contains $512^3$ dark matter and $512^3$ gas particles in a periodic box of $(100$ h$^{-1}$ $\rm {Mpc})^3$. It is referred as the `B100-SF1' run in the rest of the paper, which adopts a Plummer equivalent softening length as $\epsilon_{{\rm Pl}} = 9.8$ h$^{-1}$ kpc.
The second simulation, referred as `B100-SF2', inherits most of the parameters of the B100-SF1 run but with a different softening length $\epsilon_{{\rm Pl}} = 4.5$ h$^{-1}$ kpc to examine the possible influence from the resolution. 

\begin{figure*}
\includegraphics[width=190mm]{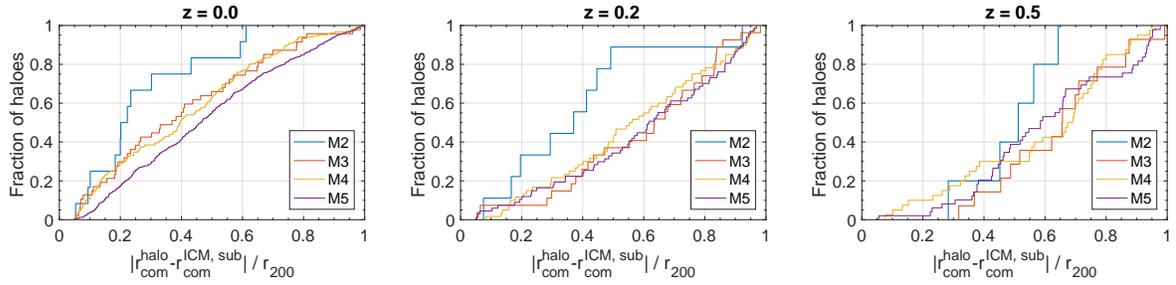}
\caption{Cumulative distributions of the rescaled offsets $r^{{\rm h,Is}}_{{\rm off}}/r_{200}$ for different mass bins in the B100-SF1 run. The offsets distance is rescaled by virial radius $r_{200}$. Lines of different colors refer to different halo mass bins as detailed in Table \ref{tab1}.}
\label{fig1}
\end{figure*}

\begin{figure*}
\includegraphics[width=190mm]{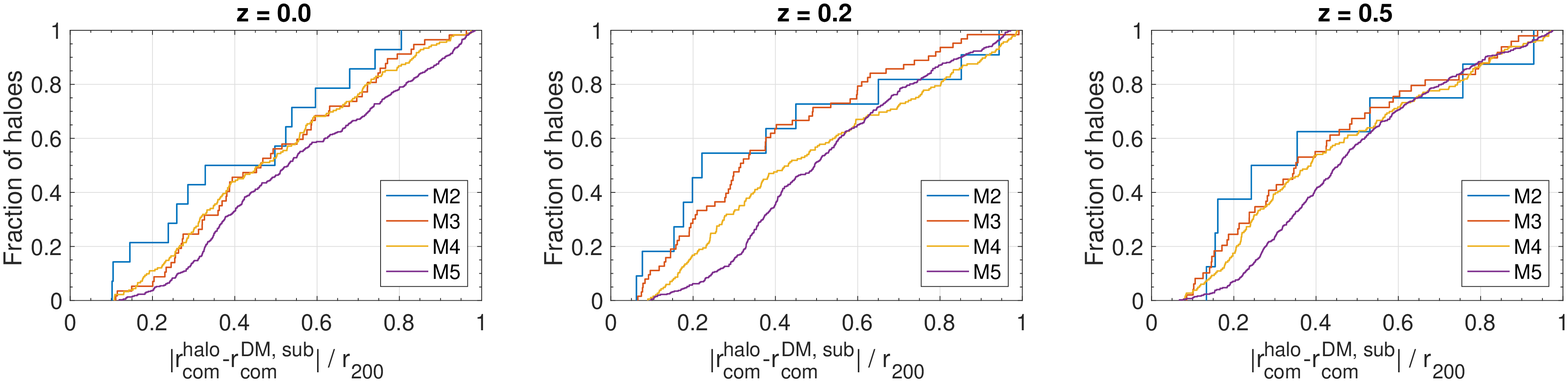}
\caption{Cumulative distributions of the rescaled offsets $r^{{\rm h,Ds}}_{{\rm off}}/r_{200}$ for different mass bins in the B100-SF1 run. The offsets distance is rescaled by virial radius $r_{200}$. Lines of different colors refer to different halo mass bins as detailed in Table \ref{tab1}.}
\label{fig2}
\end{figure*}

\begin{table*}
\begin{tabular}{|c|c|c|c|c|c|c|}
\hline
 & \multicolumn{3}{|c|}{$512^3$ B100-SF1} & \multicolumn{3}{|c|}{$512^3$ B100-SF2}\\ 
& $z=0$ & $z=0.2$ & $z=0.5$ & $z=0$ & $z=0.2$ & $z=0.5$ \\ 
\hline
{\bf{M1}}: $M_{200} \geq 10^{14.5}{\rm ~h}^{-1}~\textmd{M}_\odot$ & 3(2) & 1(1) & 1(1) &3(2) & 1(1) & 1(1) \\
{\bf{M2}}: $10^{14.0}{\rm ~h}^{-1}~\textmd{M}_\odot  \leq M_{200} < 10^{14.5}{\rm ~h}^{-1}~\textmd{M}_\odot $ & 15(12) & 12(9) & 8(5)  & 15(6) & 12(7) & 8(4) \\
{\bf{M3}}: $10^{13.5}{\rm ~h}^{-1}~\textmd{M}_\odot  \leq M_{200} < 10^{14.0}{\rm ~h}^{-1}~\textmd{M}_\odot $ & 63(47) & 66(27) & 52(14) & 63(45) & 65(30) & 55(21) \\
{\bf{M4}}: $10^{13.0}{\rm ~h}^{-1}~\textmd{M}_\odot  \leq M_{200} < 10^{13.5}{\rm ~h}^{-1}~\textmd{M}_\odot $ & 259(187) & 232(60) & 217(40) & 259(175) & 241(67) & 220(32)\\
{\bf{M5}}: $10^{12.5}{\rm ~h}^{-1}~\textmd{M}_\odot  \leq M_{200} < 10^{13.0}{\rm ~h}^{-1}~\textmd{M}_\odot $ & 800(411) & 803(67) &750(49) & 807(416) & 812(81) & 768(54) \\
\hline
\end{tabular}
\caption{Number of halos in different mass bins for the two cosmological simulations. Numbers in parentheses are the numbers of halos with $r^{{\rm h,Is}}_{{\rm off}}/r_{200}\gtrsim 0.05$.}
\label{tab1}
\end{table*}

\begin{figure*}
\includegraphics[width=160mm]{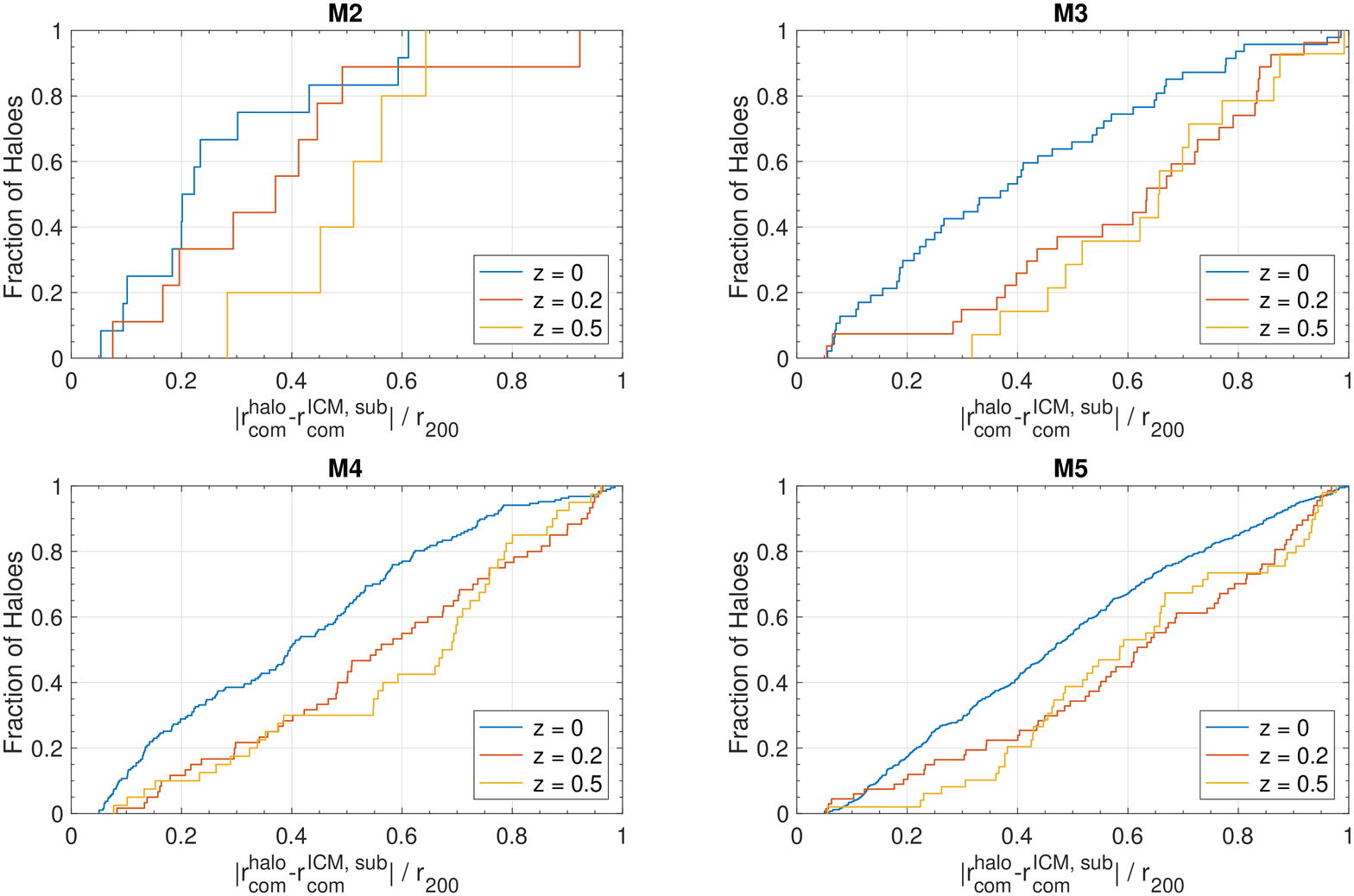}
\caption{Redshift evolution of the cumulative distributions of the rescaled offsets $r^{{\rm h,Is}}_{{\rm off}}/r_{200}$ for different mass bins in the B100-SF1 run. Lines of different colors refer to the results for different redshifts.}
\label{fig3}
\end{figure*}

\begin{figure*}
\includegraphics[width=160mm]{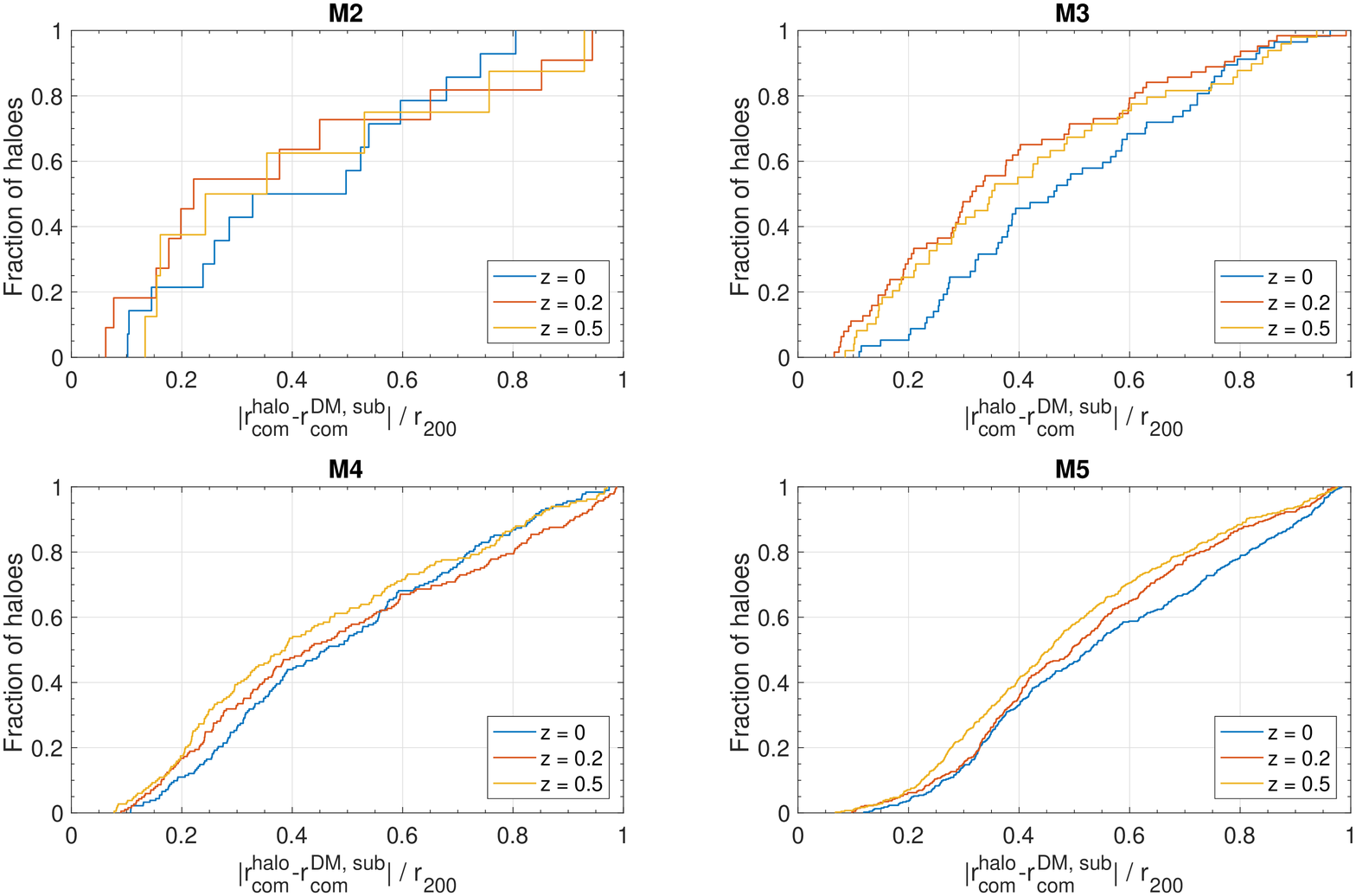}
\caption{Redshift evolution of the cumulative distributions of the rescaled offsets $r^{{\rm h,Ds}}_{{\rm off}}/r_{200}$ for different mass bins in the B100-SF1 run. Lines of different colors refer to the results for different redshifts.}
\label{fig4}
\end{figure*}

A flat $\Lambda$CDM cosmology is adopted for both simulations, where the concordance parameters take the values $\Omega_{\rm m,0} = 0.30$, $\Omega_{\rm \Lambda, 0} = 0.70$, $\Omega_{\rm b,0} = 0.045$ for the matter, the dark energy, and the baryonic matter density parameter respectively, $h = 0.70$ for the current dimensionless Hubble parameter, and $n_{\rm s} =0.96$ for the spectrum index of primordial perturbation \citep{Planck2015}. Baryonic physics such as an ultraviolet (UV) background, radiative cooling, star formation, and the feedback from supernovae are included in the simulations. Feedback from AGN is ignored at the current stage. 

The Amiga's Halo Finder (AHF, \citet{KK2009}) is used to obtain the halo catalogue given that the mean density is $\rho(<r_{200}) = 200 \rho_{\rm crit}$ within virial radius, where $\rho_{{\rm crit}}\equiv 3H_0^2/(8\pi G)$ is the critical density of the universe. Only the halos with a number of particles larger than $1000$ are considered in our study.

\section{Barycentre Offsets Between Primary and Subhaloes}
The centre-of-mass, which also called the barycentre, is one of the most common definitions of halo centre and can be easily computed from the simulation data. It is also more closely related to the inertia tensor in dark halo shape measurements \citep{Kiessling2016}. In this work, we study the three-dimension (3D) barycentre offsets between the baryons and the entire halo.

When we refer to the baryonic centre of clusters of galaxies in this study, we focus on the ICM gas component. It contributes $\sim 90$ per cent of the total baryonic mass of the entire system. For each halo, we rerun the halo finder algorithm for the gas within the virial radius to generate the catalogue of the gas subhalo. The position (centre-of-mass) of the gas subhalo with largest overdensity is dubbed as $\mathbf{r}^{\rm ICM, sub}_{\rm com}$. 
The barycentre of the entire halo (including dark matter, gas and stellar components) is denoted as $\mathbf{r}^{\rm halo}_{{\rm com}}$.  The 3D barycentre offsets between the entire halo and the gas subhalo with largest overdensity is given as
\begin{center}
\begin{equation}
r^{{\rm h,Is}}_{{\rm off}}= |\mathbf{r}^{\rm halo}_{{\rm com}} - \mathbf{r}^{\rm ICM, sub}_{{\rm com}}|.
\label{roff}
\end{equation}
\end{center}
This is a physically intrinsic offsets between the gas and the entire system and likely can be related to the physical properties of groups and clusters. 
For comparison, the barycentre offsets between the entire halo and the dark matter subhalo with largest overdensity is also calculated, i.e.
\begin{center}
\begin{equation}
r^{{\rm h,Ds}}_{{\rm off}}= |\mathbf{r}^{\rm halo}_{{\rm com}} - \mathbf{r}^{\rm DM, sub}_{{\rm com}}|.
\label{roff2}
\end{equation}
\end{center}
$\mathbf{r}^{\rm DM, sub}_{{\rm com}}$ is the position (centre-of-mass) of the dark matter subhalo with largest overdensity. It is obtained in a similar way as $\mathbf{r}^{\rm ICM, sub}_{{\rm com}}$ but for the dark matter within the virial radius fo the system.

We perform our statistical analysis for three different snapshots (different redshifts), i.e. at $z= 0, 0.2$ and $0.5$ to examine the possible redshift evolution of the results. The halos at $z= 0.2$ and $0.5$ are identified by using the same spherical overdensity method as those at $z= 0$. In each snapshot, we divide the halo catalogue into five different mass bins, with a bin width of $\Delta {\rm log}(M_{200}/{\rm ~h}^{-1}~\textmd{M}_\odot ) = 0.5$. The number of halos in different mass and redshift bins are given in Table \ref{tab1}. We focus on those halos with $r^{{\rm h,Is}}_{{\rm off}}/r_{200}, r^{{\rm h,Ds}}_{{\rm off}}/r_{200}\gtrsim 0.05$, where $r_{200}$ is the virial radius of the halo.

The cumulative distribution functions (hereafter CDFs) of $r^{{\rm h,Is}}_{{\rm off}}/r_{200}$ and $r^{{\rm h,Ds}}_{{\rm off}}/r_{200}$ for the halo catalogue in the B100-SF1 run are shown respectively in Figure \ref{fig1} and \ref{fig2}. 
In both figures, one can see that the halos in a more massive bins are likely to have a larger proportion of halos with smaller offsets (i.e. a steeper CDF), although this trend is more apparently manifested in the M2 bin compared to the rest mass bins. The halos in the M2 bin tend to have relatively smaller $r^{{\rm h,Is}}_{{\rm off}}/r_{200}$ as well as $r^{{\rm h,Ds}}_{{\rm off}}/r_{200}$ than those halos with smaller mass $M_{200}$.
This can possibly be explained by the fact that smaller halos usually have larger merger rates than those more massive ones so that the former are often found to have larger offsets. 
To be concise, we mainly present the results from the fiducial B100-SF1 run, given that the results from the B100-SF2 run are consistent with those obtained from the B100-SF1 run. The results from the B100-SF2 run are shown wherever necessary.

The redshift evolution of the CDF of $r^{{\rm h,Is}}_{{\rm off}}/r_{200}$ and $r^{{\rm h,Ds}}_{{\rm off}}/r_{200}$ are presented respectively in Figure \ref{fig3} and \ref{fig4}. For each mass bin, the proportion of halos with a small offsets $r^{{\rm h,Is}}_{{\rm off}}/r_{200}$ is larger for lower redshifts. It implies that the halos at lower redshifts are likely to have smaller offsets $r^{{\rm h,Is}}_{{\rm off}}/r_{200}$ than those at higher redshifts. This is in agreement with the current structure formation theory. In the $\Lambda$CDM universe, the merge frequency slows down in the redshift range considered here, and hence the relaxed-ness increases along with time. The above features and explanations also hold their validity for the CDFs of $r^{{\rm h,Ds}}_{{\rm off}}/r_{200}$ in Figure \ref{fig4}.
The result for the mass bin M1 is contaminated by the statistical noise due to the deficiency of halo samples and is therefore not shown in Figure \ref{fig1} to \ref{fig4}.

\begin{figure*}
\includegraphics[width=190mm]{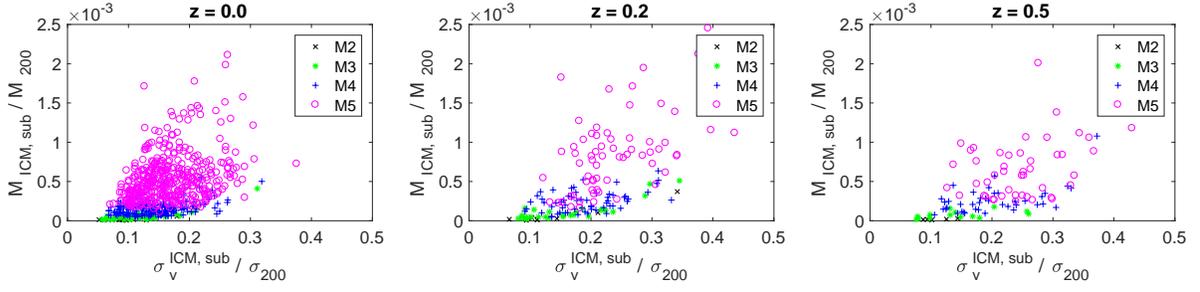}
\caption{Mass as a function of the velocity dispersion $\sigma_{v}^{\rm ICM,sub}/\sigma_{200}$ for the gas subhalo with maximum overdensity in the B100-SF1 run. $M_{\rm ICM, sub}$ represents the mass of the gas subhalo and $M_{\rm 200}$ is the mass of the whole system. The gas velocity dispersion $\sigma_{v}^{\rm ICM,sub}$ is rescaled by the velocity dispersion $\sigma_{200}$ for the whole system, which is calculated from equation (\ref{sigma200}). Data points with different symbols represent halos in the five different halo mass bins as detailed in Table \ref{tab1}. The results are presented at three different redshifts $z= 0, 0.2$, and $0.5$.}
\label{fig5}
\end{figure*}

\begin{figure*}
\includegraphics[width=190mm]{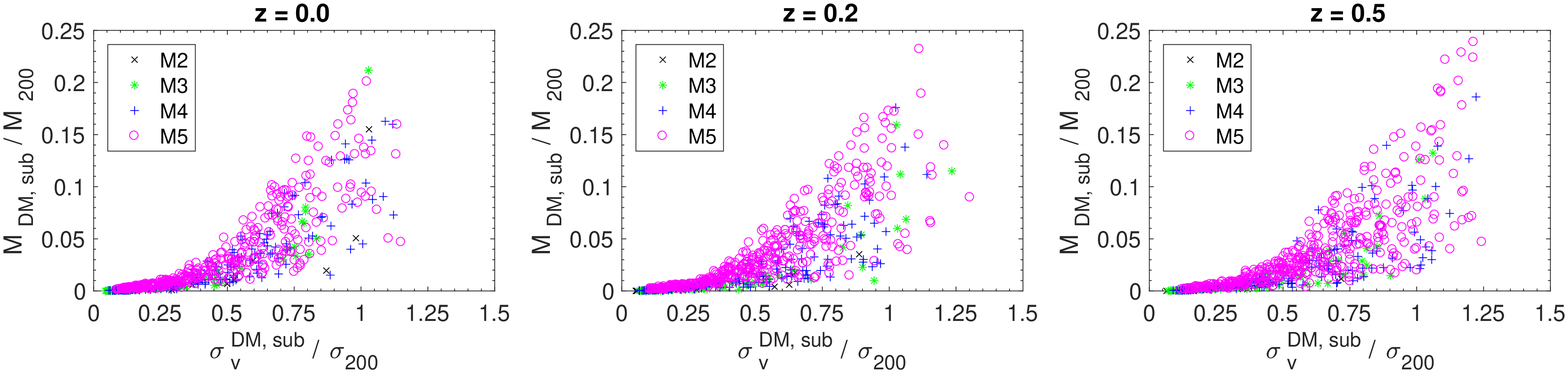}
\caption{Mass as a function of the velocity dispersion $\sigma_{v}^{\rm DM,sub}/\sigma_{200}$ for the dark matter subhalo with maximum overdensity in the B100-SF1 run. $M_{\rm ICM, sub}$ represents the mass of the dark matter subhalo and $M_{\rm 200}$ is the mass of the whole system. The dark matter velocity dispersion $\sigma_{v}^{\rm DM,sub}$ is rescaled by the velocity dispersion $\sigma_{200}$ for the whole system, which is calculated from equation (\ref{sigma200}). Data points with different symbols represent halos in the five different halo mass bins as detailed in Table \ref{tab1}. The results are presented at three different redshifts $z= 0, 0.2$, and $0.5$.}
\label{fig6}
\end{figure*}

We also study the relation between the velocity dispersion and the mass fraction for both of the gas and dark matter subhalo. In Figure \ref{fig5}, we present the fraction of gas in the subhalo as a function of the 3D velocity dispersion of the gas component for the B100-SF1 run. Like the rescaled offsets, we introduce the rescaled velocity dispersion $\sigma_{v}^{\rm ICM,sub}/\sigma_{200}$ to reduce the possible bias induced by the mass of the system, where the gas velocity dispersion $\sigma_{v}^{\rm ICM,sub}$ of the maximum-overdensity gas subhalo is rescaled by the 3D velocity dispersion $\sigma_{200}$ of the entire system. According to \citet{Munari2013}, $\sigma_{200}$ is given as
\be
\frac{\sigma_{200}}{{\rm km~s}^{-1}}=\frac{\sqrt{3}\sigma_{\rm 1D}}{{\rm km~s}^{-1}}=\sqrt{3} A_{\rm 1D}\left[\frac{h(z) M_{200}}{10^{15.0}\textmd{M}_\odot}\right]^{\alpha}, 
\label{sigma200}
\ee
where $A_{\rm 1D}\simeq 10^3$ and $\alpha \simeq 1/3$. $\sigma_{200}$ and $M_{200}$ are in unit of km s$^{-1}$ and h$^{-1}$ $\textmd{M}_\odot$ respectively. 
Similar to Figure \ref{fig5}, in Figure \ref{fig6} we present the mass fraction of dark matter in the subhalo with largest overdensity of dark matter as a function of the 3D velocity dispersion of the dark matter component for the B100-SF1 run.

Two comments are necessary for both Figure \ref{fig5} and \ref{fig6}.
One is that it shows that for the unrelaxed halos we study in this work, the gas subhalo with maximum overdensity and larger gas mass fraction $M_{\rm ICM, sub}/M_{200}$ tend to have larger rescaled gas velocity dispersion $\sigma_{v}^{\rm ICM,sub}/\sigma_{200}$. 
A larger rescaled dark matter velocity dispersion $\sigma_{v}^{\rm DM,sub}/\sigma_{200}$ is also found for the dark matter subhalo with maximum overdensity and larger mass fraction $M_{\rm DM, sub}/M_{200}$ of dark matter.
For the gas subhalos, this correlation is more apparent for those halos in the snapshot $z=0$. The unrelaxed halos in the snapshots $z=0.2$ and $z=0.5$ are more scattered and the correlation between the gas mass fraction $M_{\rm ICM, sub}/M_{200}$ and the gas velocity dispersion $\sigma_{v}^{\rm ICM,sub}/\sigma_{200}$ is weak. 
The other is that the correlation between the rescaled mass $M_{\rm ICM, sub}/M_{200}$ and the gas velocity dispersion $\sigma_{v}^{\rm ICM,sub}/\sigma_{200}$ as shown in Figure \ref{fig5} (as well as the correlation between the rescaled mass $M_{\rm DM, sub}/M_{200}$ and the gas velocity dispersion $\sigma_{v}^{\rm DM,sub}/\sigma_{200}$ as shown in Figure \ref{fig6}) is similar to the $M_{\rm BH}-\sigma$ relation for galaxies. The $M_{\rm BH}-\sigma$ relation is an empirical correlation between the 3D velocity dispersion $\sigma$ of the stellar components in a galaxy bulge and the mass $M_{\rm BH}$ of its centre black hole. It is believed to be related to some unknown feedback mechanism between the formation of galaxy bulges and the growth of centre black holes \citep{McConnell2011,King2013}. Our results imply that this relation, despite the large scatter, may possibly still hold true for both the gas and dark matter subhalos within the unrelaxed halos.

\begin{figure*}
\includegraphics[width=190mm]{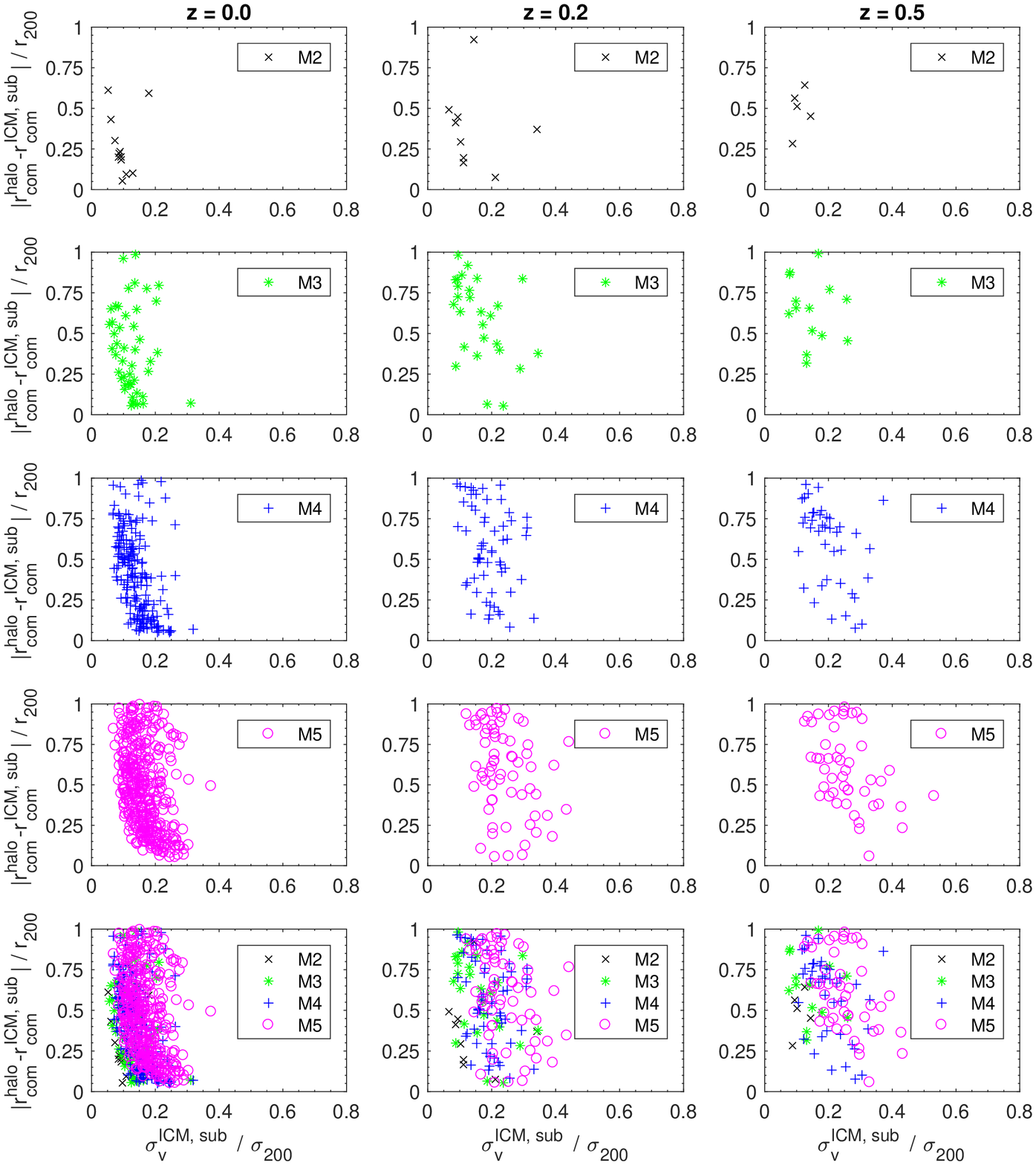}
\caption{Rescaled offsets, $r^{{\rm h,Is}}_{{\rm off}}/r_{200}$ as a function of the rescaled gas velocity dispersion, $\sigma_{v}^{\rm ICM,sub}/\sigma_{200}$ of the highest overdensity dark matter subhalo for the B100-SF1 run. Results for different mass bins at three different redshifts $z= 0, 0.2$, and $0.5$ are presented.
}
\label{fig7}
\end{figure*}

We are also interested in the correlation between the rescaled spatial offsets $r^{{\rm h,Is}}_{{\rm off}}/r_{200}$ or $r^{{\rm h,Ds}}_{{\rm off}}/r_{200}$ and the 3D velocity dispersion of the subhalo with the maximum overdensity. We pay particular attention to those unrelaxed halos which have large spatial offsets, i.e. $r^{{\rm h,Is}}_{{\rm off}}/r_{200}, r^{{\rm h,Ds}}_{{\rm off}}/r_{200}\gtrsim 0.05$. In Figure \ref{fig7}, we plot the $r^{{\rm h,Is}}_{{\rm off}}/r_{200}$ as a function of $\sigma_{v}^{\rm ICM,sub}/\sigma_{200}$ for halos in different mass bins and redshifts.
One can see that there is an obvious anti-correlation between the rescaled barycentre offsets $r^{{\rm h,Is}}_{{\rm off}}/r_{200}$ and gas velocity dispersion $\sigma_{v}^{\rm ICM,sub}/\sigma_{200}$ for the unrelaxed halos in the snapshot $z=0$. In each mass bin, the gas subhalos with maximum overdensity and smaller velocity dispersion $\sigma_{v}^{\rm ICM,sub}/\sigma_{200}$ tend to have larger rescaled barycentre offsets $r^{{\rm h,Is}}_{{\rm off}}/r_{200}$. This feature can also be observed from the halos in the snapshot $z=0.2$ and $z=0.5$ but with larger scatter. The mass bin M1 is insufficiently sampled and is therefore not presented and ignored in our discussion.

\begin{figure*}
\includegraphics[width=190mm]{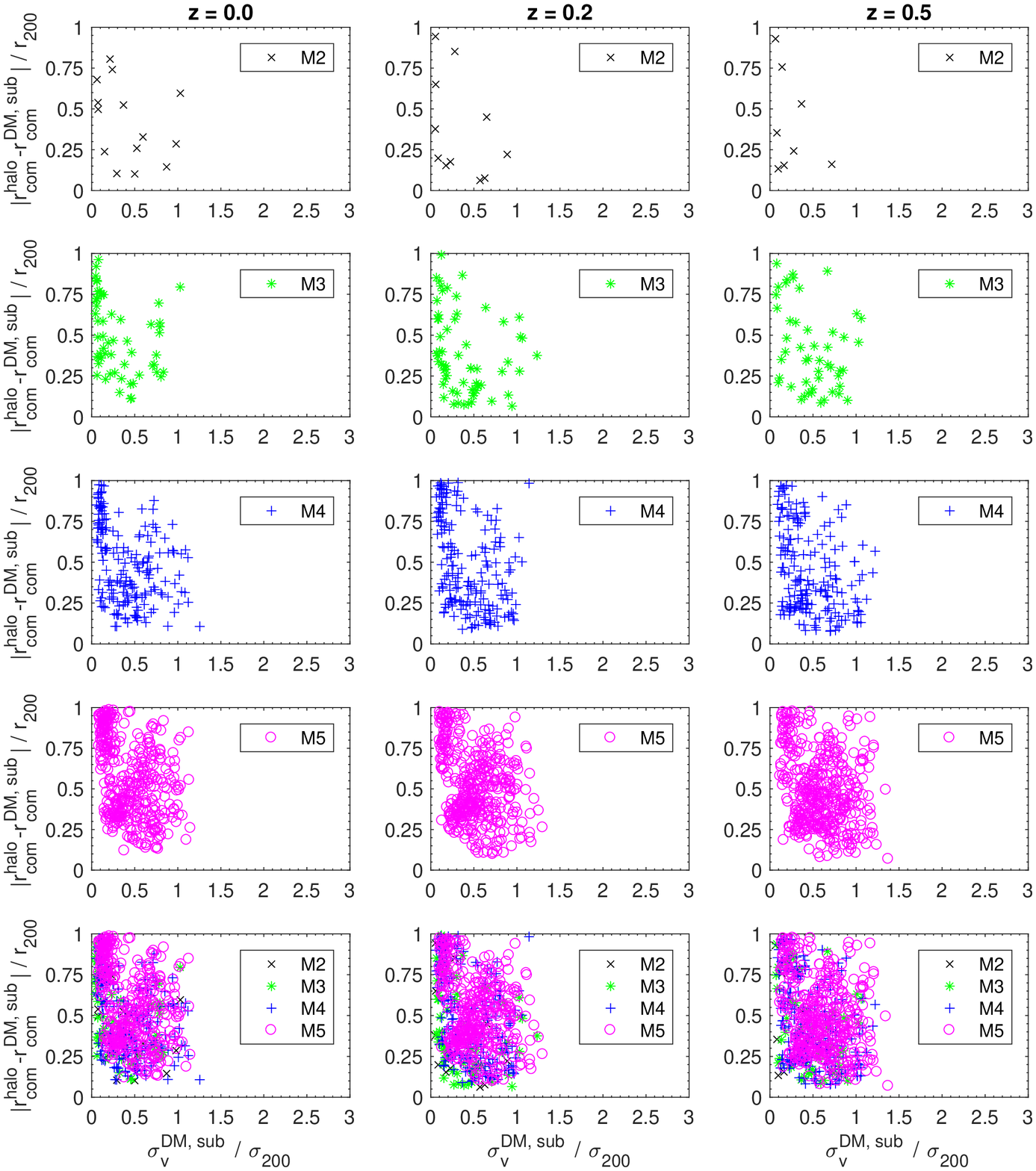}
\caption{Rescaled offsets, $r^{{\rm h,Ds}}_{{\rm off}}/r_{200}$ as a function of the rescaled velocity dispersion, $\sigma_{v}^{\rm DM, sub}/\sigma_{200}$ of the highest overdensity dark matter subhalo for the B100-SF1 run. Results for different mass bins at three different redshifts $z= 0, 0.2$, and $0.5$ are presented.
}
\label{fig8}
\end{figure*}

The $r^{{\rm h,Ds}}_{{\rm off}}/r_{200}$ as a function of $\sigma_{v}^{\rm DM,sub}/\sigma_{200}$ for halos in different mass bins and redshifts are plotted in Figure \ref{fig8}. Similar anti-correlation between $r^{{\rm h,Ds}}_{{\rm off}}/r_{200}$ and the velocity dispersion $\sigma_{v}^{\rm DM,sub}/\sigma_{200}$ for the highest overdensity dark matter subhalos have been found but with less significance than that for $r^{{\rm h,Is}}_{{\rm off}}/r_{200}$ to $\sigma_{v}^{\rm ICM,sub}/\sigma_{200}$ in Figure \ref{fig7}. Halos with large offsets $r^{{\rm h,Ds}}_{{\rm off}}/r_{200}$ are likely to harbor highest overdensity dark matter subhalos with relatively small velocity dispersion $\sigma_{v}^{\rm DM,sub}/\sigma_{200}$. It should also be noticed that the above feature holds true for the snapshot $z=0, 0.2$ as well as $z=0.5$, while for the gas subhalos, this phenomenon is more significant in the snapshot $z=0$ than $z=0.2, 0.5$.

\begin{figure*}
\includegraphics[width=190mm]{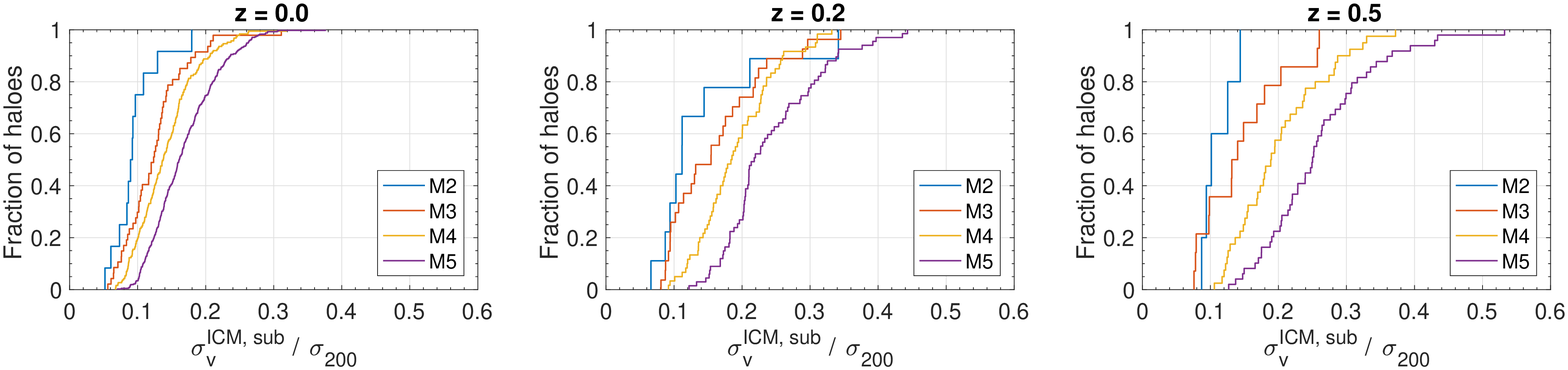}
\caption{Cumulative distributions of the rescaled offsets $\sigma_{v}^{\rm ICM, sub}/\sigma_{200}$ for different mass bins in the B100-SF1 run. The gas velocity dispersion $\sigma_{v}^{\rm ICM,sub}$ is rescaled by the velocity dispersion $\sigma_{200}$ for the whole system. The results are presented at three different redshifts $z= 0, 0.2$, and $0.5$. Lines of different colors refer to different halo mass bins as detailed in Table \ref{tab1}.}
\label{fig9}
\end{figure*}

\begin{figure*}
\includegraphics[width=190mm]{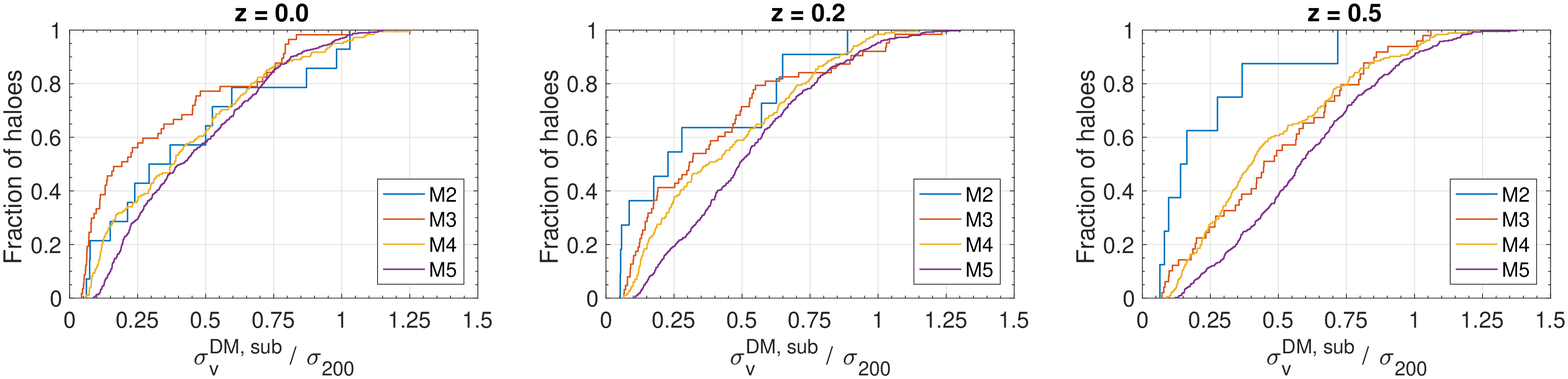}
\caption{Cumulative distributions of the rescaled offsets $\sigma_{v}^{\rm DM, sub}/\sigma_{200}$ for different mass bins in the B100-SF1 run. The velocity dispersion of dark subhalo $\sigma_{v}^{\rm DM,sub}$ is rescaled by the velocity dispersion $\sigma_{200}$ for the whole system. The results are presented at three different redshifts $z= 0, 0.2$, and $0.5$. Lines of different colors refer to different halo mass bins as detailed in Table \ref{tab1}.}
\label{fig10}
\end{figure*}

To find out the cause of this anti-correlation, we take a closer look at the 3D velocity dispersion of the gas subhalo with the maximum overdensity. The CDF of $\sigma_{v}^{\rm ICM,sub}/\sigma_{200}$ and $\sigma_{v}^{\rm DM,sub}/\sigma_{200}$ for halos in different mass bins and redshifts are respectively plotted in Figure \ref{fig9} and \ref{fig10}. In Figure \ref{fig9}, one can see that the gas velocity dispersion $\sigma_{v}^{\rm ICM,sub}/\sigma_{200}$ for the gas subhalo shows evident dependence on the halo mass $M_{200}$. The halos resides in the more massive bins, i.e. the M2 and M3 bin, are likely to contain the gas subhalos which has smaller $\sigma_{v}^{\rm ICM,sub}/\sigma_{200}$ and maximum overdensity. 
The same feature has also been found in Figure \ref{fig10} for the CDF of $\sigma_{v}^{\rm DM,sub}/\sigma_{200}$.
These are in coincidence with what we find in Figure \ref{fig1} and \ref{fig2} that the massive halos usually have lower merge frequency than those less massive ones so that the former usually have smaller offsets $r^{{\rm h,Is}}_{{\rm off}}/r_{200}$ and $r^{{\rm h,Ds}}_{{\rm off}}/r_{200}$. This explanation may also help to account for the above features of both the $\sigma_{v}^{\rm ICM,sub}/\sigma_{200}$ and $\sigma_{v}^{\rm DM,sub}/\sigma_{200}$.

\begin{figure*}
\includegraphics[width=160mm]{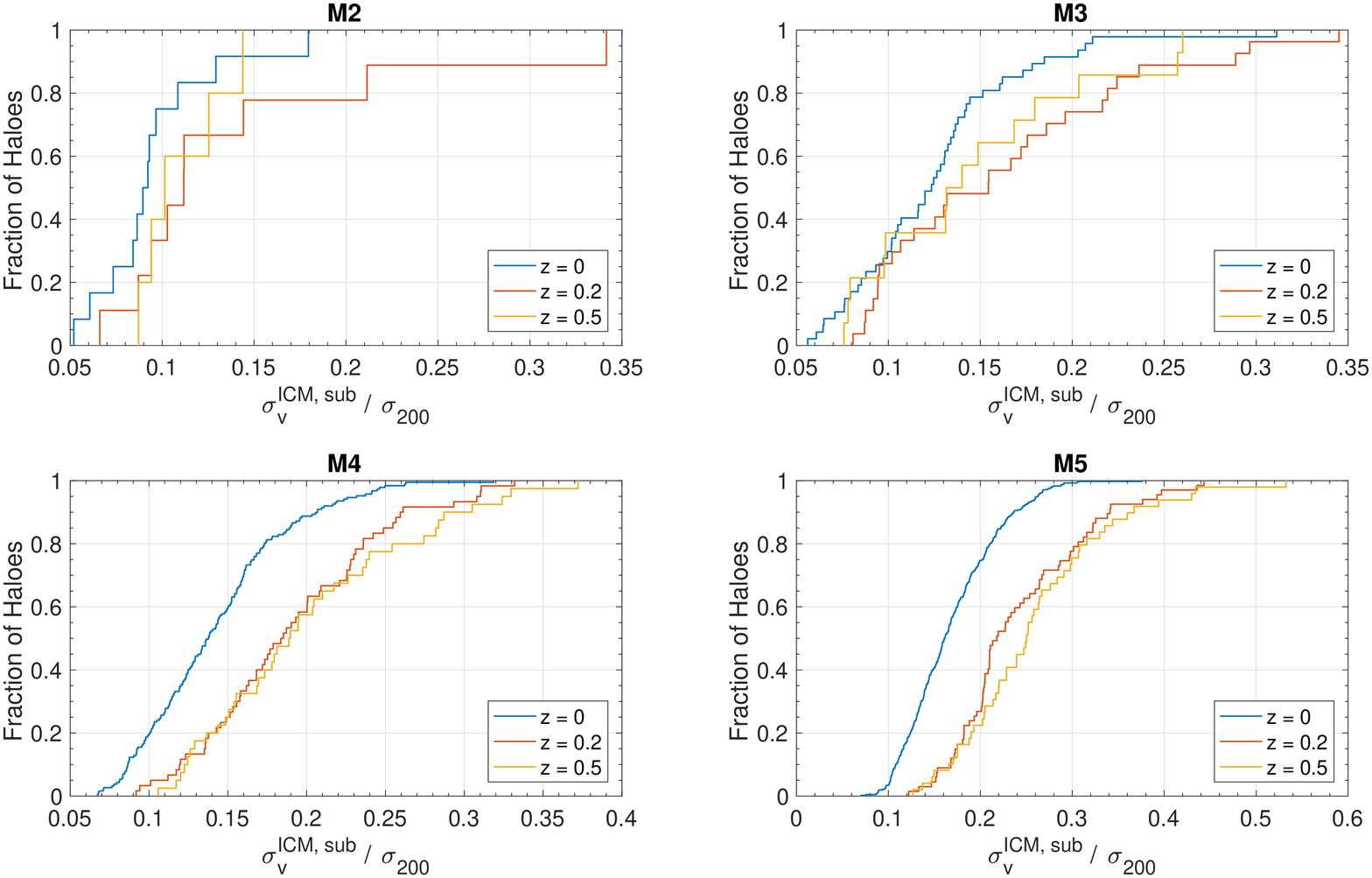}
\caption{Redshift evolution of the cumulative distributions of the rescaled offsets $\sigma_{v}^{\rm ICM,sub}/\sigma_{200}$ for different mass bins in the B100-SF1 run. The gas velocity dispersion $\sigma_{v}^{\rm ICM,sub}$ is rescaled by the velocity dispersion $\sigma_{200}$ for the whole system. Lines of different colors refer to the results from three different redshifts $z= 0, 0.2$, and $0.5$.}
\label{fig11}
\end{figure*}

\begin{figure*}
\includegraphics[width=160mm]{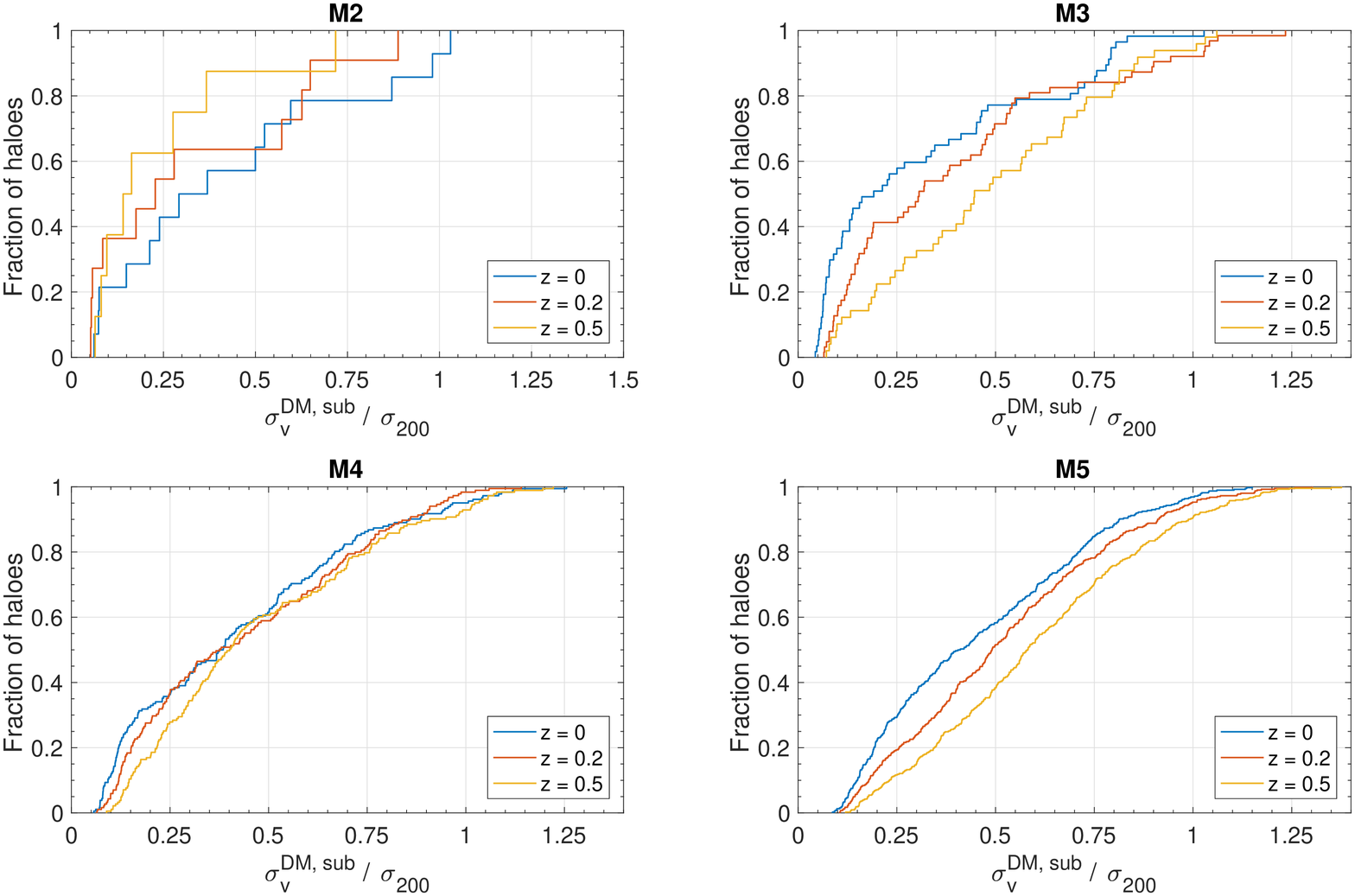}
\caption{Redshift evolution of the cumulative distributions of the rescaled offsets $\sigma_{v}^{\rm DM,sub}/\sigma_{200}$ for different mass bins in the B100-SF1 run. The dark matter velocity dispersion $\sigma_{v}^{\rm DM,sub}$ is rescaled by the velocity dispersion $\sigma_{200}$ for the whole system. Lines of different colors refer to the results from three different redshifts $z= 0, 0.2$, and $0.5$.}
\label{fig12}
\end{figure*}

In Figure \ref{fig11} and \ref{fig12}, we respectively present the redshift evolution of the CDF of $\sigma_{v}^{\rm ICM,sub}/\sigma_{200}$ and $\sigma_{v}^{\rm DM,sub}/\sigma_{200}$. Like the case of $r^{{\rm h,Is}}_{{\rm off}}/r_{200}$ in Figure \ref{fig3}, the gas contained in halos with higher redshifts tend to have larger velocity dispersion $\sigma_{v}^{\rm ICM,sub}/\sigma_{200}$ as well as larger offsets $r^{{\rm h,Is}}_{{\rm off}}/r_{200}$. This also holds true for the maximum-overdensity dark matter subhalo. As shown in Figure \ref{fig12} and \ref{fig3}, the dark matter subhalos contained in halos with higher redshifts are likely to have larger velocity dispersion $\sigma_{v}^{\rm DM,sub}/\sigma_{200}$ and larger offsets $r^{{\rm h,Ds}}_{{\rm off}}/r_{200}$. All these facts are consistent with what we expect from the current structure formation theory that the un-relaxedness of the halos increases along with redshifts.

\section{Conclusions and Discussion}
In this paper, we used two hydrodynamic simulations to study the barycentre offsets $r_{{\rm off}}^{{\rm h,Is}}/r_{200}$ between the entire halo and the gas subhalo with the maximum overdensity. The latter is closely related to the most condense baryonic region of the ICM gas in clusters and groups of galaxies. 
For comparison, we repeated the above analysis on the dark matter substructure as well to obtain the barycentre offsets $r_{{\rm off}}^{{\rm h,Ds}}/r_{200}$ between the entire halo and the dark matter subhalo with the highest overdensity.
We restricted our analysis to those halos with large offsets, i.e. $r_{{\rm off}}^{{\rm ICM,sub}}/r_{200}, r_{{\rm off}}^{{\rm h,Ds}}/r_{200}\gtrsim 0.05$. We found that the halos at higher redshifts are more likely to have larger $r_{{\rm off}}^{{\rm ICM,sub}}/r_{200}$ and $r_{{\rm off}}^{{\rm h,Ds}}/r_{200}$, which is consistent with what we expect from the current structure formation theory in the $\Lambda$CDM universe that the merge rate of galaxies and clusters of galaxies goes down in the redshift range we studied here. The 3D velocity dispersion of the gas and dark matter subhalos with maximum overdensity have a similar mass and redshift dependence as $r_{{\rm off}}^{{\rm ICM,sub}}/r_{200}$ and $r_{{\rm off}}^{{\rm h,Ds}}/r_{200}$. The gas and dark matter subhalos with maximum overdensity and at lower redshifts usually have smaller $\sigma_{v}^{\rm ICM,sub}/\sigma_{200}$ and $\sigma_{v}^{\rm DM,sub}/\sigma_{200}$. For the unrelaxed halos which have rescaled offsets $r_{{\rm off}}^{{\rm h,Is}}/r_{200}\gtrsim 0.05$, we found that the gas subhalos with maximum overdensity and larger gas velocity dispersion $\sigma_{v}^{\rm ICM,sub}/\sigma_{200}$ tend to have smaller barycentre offsets $r_{{\rm off}}^{{\rm 3D}}/r_{200}$ between the gas subhalo and the entire halo, as shown in Figure \ref{fig7}. Similar anti-correlation between the velocity dispersion $\sigma_{v}^{\rm DM,sub}/\sigma_{200}$ of the largest-overdensity dark matter subhalo and $r_{{\rm off}}^{{\rm h,Ds}}/r_{200}$ was also found in Figure \ref{fig8} although it was less significant.

It should be noted that when we calculated the spatial offsets $r_{{\rm off}}^{{\rm h,Is}}/r_{200}, r_{{\rm off}}^{{\rm h,Ds}}/r_{200}$ and the velocity dispersion $\sigma_{v}^{\rm ICM,sub}/\sigma_{200}, \sigma_{v}^{\rm DM,sub}/\sigma_{200}$, we focused on the subhalo which has highest overdensity in the cluster. The primary gas or dark matter halo was ignored in our analysis in this paper. In fact, the spatial offsets between the primary gas halo and the entire cluster have been studied in another paper \citep{Li2018}. In \citet{Li2018}, the centre offsets between the primary ICM gas halo and the entire cluster were studied using three hydrodynamical cosmological simulations. 
Correlation between the offsets and the 3D velocity dispersion of the overall gas halo was found. With respect to the centre offsets in clusters of galaxies, our work in this paper showed that the results for the subhalos are quite different from that for the primary halos (refer to \citet{Li2018} for more details).

The anti-correlation between $r_{{\rm off}}^{{\rm h,Is}}/r_{200}, r_{{\rm off}}^{{\rm h,Ds}}/r_{200}$ and $\sigma_{v}^{\rm ICM,sub}/\sigma_{200}, \sigma_{v}^{\rm DM,sub}/\sigma_{200}$ for the largest overdensity subhalos we discovered in this paper are in contradiction with what we expect from the current structure formation theory. In the current theory of structure formation, the barycentre offsets $r_{{\rm off}}^{{\rm h,Is}}/r_{200}$ or $r_{{\rm off}}^{{\rm h,Ds}}/r_{200}$ can be taken as indicator of the dynamical state of the halo system. The halos with large $r_{{\rm off}}^{{\rm h,Is}}/r_{200}, r_{{\rm off}}^{{\rm h,Ds}}/r_{200}$ are more likely to be dynamically unrelaxed than those with smaller $r_{{\rm off}}^{{\rm h,Is}}/r_{200}, r_{{\rm off}}^{{\rm h,Ds}}/r_{200}$. A larger 3D velocity dispersion should be expected for these halos. One would intuitively expect a larger $\sigma_{v}^{\rm ICM,sub}/\sigma_{200}, \sigma_{v}^{\rm DM,sub}/\sigma_{200}$ for the largest overdensity subhalos. However, Figure \ref{fig7} and \ref{fig8} show another picture. The subhalos with maximum overdensity and larger velocity dispersion $\sigma_{v}^{\rm ICM,sub}/\sigma_{200}, \sigma_{v}^{\rm DM,sub}/\sigma_{200}$ tend to reside in those halos with smaller barycentre offsets $r_{{\rm off}}^{{\rm h,Is}}/r_{200}, r_{{\rm off}}^{{\rm h,Ds}}/r_{200}$ (equivalently more dynamically relaxed). 

Although the two separate simulations in our study both demonstrate the existence of such an anti-correlation, the underlying physical mechanism remains unclear. One can obtain from the virial theorem that $r_{200}\sim GM_{200}/{(\sigma_v)}^2$ and naturally assumes that the spatial offsets $r_{\rm off}\sim r_{200}\sim GM_{200}/{(\sigma_v)}^2$. But the large $r_{\rm off}/r_{200}$ for the halos in our analysis suggests that these halos are probably dynamically unrelaxed and do not satisfy the virial theorem. It should also be noticed that in Figure \ref{fig7}, the evidence of such an anti-correlation between $r_{{\rm off}}^{{\rm h,Is}}/r_{200}$ and $\sigma_{v}^{\rm ICM,sub}$ is more significant for the snapshot $z=0$ than $z=0.2$ and $z=0.5$. 
This feature was not observed for the $r_{{\rm off}}^{{\rm h,Ds}}/r_{200}-\sigma_{v}^{\rm DM,sub}$ anti-correlation in Figure \ref{fig8}. It implies that the physical mechanisms behind these two similar anti-correlations are probably different. For the highest-overdensity gas subhalos,
one may suggest that the anti-correlation can possibly be caused by the peculiar motion of the baryonic matter inside the halo. The Hubble flow of the matter caused by the cosmic expansion could `dilute' such a process so that the anti-correlation between the $r_{{\rm off}}^{{\rm h,Is}}/r_{200}$ and $\sigma_{v}^{\rm ICM,sub}$ is less significant for the snapshot at higher redshifts, i.e. $z=0.2$ and $z=0.5$. Whether this conclusion and the anti-correlation hold true for the cluster halos or halos at higher redshifts are still subject to further investigations.

A possible explanation for this anti-correlation for the gas subhalo can be found in \citet{Liao2016}. In their work, \citet{Liao2016} reported that the initial segregation between the gas and dark matter can be mainly attributed to the different physics obeyed by these two kinds of matter. During the halo merges, the gas would be stripped out by the ram pressure while the collisionless dark matter would just pass through, which leaves a gas-free halo \citep{BNA2013}.
We would take one step further to suggest that the anti-correlation between $r_{{\rm off}}^{{\rm h,Is}}/r_{200}$ and $\sigma_{v}^{\rm ICM,sub}/\sigma_{200}$ in Figure \ref{fig7} could possibly be caused by an un-head-on-head collision between two halos. In such cases, one halo sidewipes the other. Only the dark matter in the outskirt regions of each halo are gravitationally disrupted. The most condense gas region, i.e. the gas subhalo with largest overdensity, resides deep in the core of the halo and are therefore less likely to be disturbed. This could result in a small $\sigma_{v}^{\rm ICM,sub}/\sigma_{200}$ but a larger barycentre offsets $r_{{\rm off}}^{{\rm h,Is}}/r_{200}$ between the gas subhalo and the entire system. A natural conclusion from this scenario is that the less massive halos, which has smaller $M_{200}$, are more likely to be influenced by other merging halos than those more massive ones. One should expect that the less massive halos would have larger barycentre offsets $r_{{\rm off}}^{{\rm 3D}}/r_{200}$ (more disturbed by their merging parters). This is coincident with what we found from Figure \ref{fig1} (more apparent for the snapshot $z=0$). In this possible scenario, the barycentre offsets $r_{{\rm off}}^{{\rm 3D}}/r_{200}$ are mainly contributed by the distortion of dark matter, which contributes $\sim 90$ per cent of the total mass of the system.

Another possible scenario of this anti-correlation is that the large barycentre offsets $r_{{\rm off}}^{{\rm 3D}}/r_{200}$ is due to the large displacement of the gas subhalo instead of the distortion of the ourskirt dark matter in the system. In this picture, the most condense gas region is not located at the centre-of-mass for the entire halo for some other unknown mechanisms. These halos could even be dynamically relaxed, since some of them have relatively small $\sigma_{v}^{\rm ICM,sub}/\sigma_{200}$.

A possible way to test these hypotheses is to study the shape as well as the galaxy alignments in these halos. 
The underlying physical mechanisms for this phenomenon are probably not unique and are not limited to those discussed in this paper. We reported this abnormal phenomenon just to raise the possible attention of other colleagues. It deserves more systematic study and should be subject to test by more simulations with higher resolutions and baryonic processes in future.

\section*{Acknowledgments}
We would like to thank the anonymous referee for his insightful comments and constructive suggestions in improving the manuscript.
We would like to express our gratitude to Dr. Wei-Peng Lin from the Sun Yat-Sen University for providing us with the simulation data for the research.
We are grateful to the stimulating discussions with Dr. Wei-Shan Zhu, Dr. Yang Wang, and Dr. Shi-Hong Liao. 
We would like to extend our sincere gratitude to Dr. Long-Long Feng and Dr. Miao Li for their selfless assistance during the study. 
We also thank Dr. Zhi-Qi Huang, Dr. Yi-Jung Yang, and Dr. Fu-Peng Zhang for their insightful comments on this work. LMH acknowledges support from the National Natural Science Foundation of China (NSFC, Grant No. 11733010). LHN acknowledges support from the National Natural Science Foundation of China (NSFC, Grant No. 11603005).

\appendix




\begin{thebibliography}{}
\bibitem[\protect\citeauthoryear{Ade et al.}{2016}]{Planck2015}
Ade, P. A. R. et al. [Planck Collaboration] 2016, A\&A, 594, A13

\bibitem[\protect\citeauthoryear{Andreon \& Moretti}{2011}]{AM2011}
Andreon, S. \& Moretti, A. 2011, A\&A, 536, A37

 \bibitem[\protect\citeauthoryear{Brada\v{c} et al.}{2006}]{Bradac2006}
Brada\v{c} M., Clowe D., Gonzalez A. H. et al. 2006, ApJ, 652, 937

 \bibitem[\protect\citeauthoryear{Benitez-Llambay et al.}{2013}]{BNA2013}
Benitez-Llambay A., Navarro J. F., Abadi M. G., et al., 2013, ApJ, 763, L41

\bibitem[\protect\citeauthoryear{Clowe et al.}{2006}]{Clowe2006}
Clowe D., Brada\v{c} M., Gonzalez A. H., Markevitch M., Randall S. W., Jones C., Zaritsky D. 2006, ApJ, 648, L109

\bibitem[\protect\citeauthoryear{Cui et al.}{2016}]{cui2016}
Cui, Weiguang, Power C., Biffi, V., et al. 2016, MNRAS, 456, 2566

\bibitem[\protect\citeauthoryear{George et al.}{2012}]{george2012}
George M. R., Leauthaud A., Bundy K., et al. 2012, ApJ, 757, 2

\bibitem[\protect\citeauthoryear{Hilbert \& White}{2010}]{HW2010}
Hilbert, S., \& White, S. D. M., 2010, MNRAS, 404, 486

\bibitem[\protect\citeauthoryear{Hudson et al.}{2010}]{Hudson2010}
Hudson, D. S.; Mittal, R.; Reiprich, T. H.; Nulsen, P. E. J.; Andernach, H.; Sarazin, C. L., 2010, A\&A, 513, 37

 \bibitem[\protect\citeauthoryear{Johnston et al.}{2007a}]{john07a}
Johnston D. E., Sheldon E. S., Tasitsiomi A., et al. 2007a, ApJ, 656, 27

 \bibitem[\protect\citeauthoryear{Johnston et al.}{2007b}]{john07b}
Johnston D. E., Sheldon E. S., Wechsler R. H., et al. 2007b, arXiv:0709.1159

 \bibitem[\protect\citeauthoryear{Kiessling et al.}{2016}]{Kiessling2016}
Kiessling, A. et al., 2016, Space Science Reviews, 193, 139-211

 \bibitem[\protect\citeauthoryear{King}{2013}]{King2013}
King, A. 2013, ApJ, 596, L27. arXiv:astro-ph/0308342

\bibitem[\protect\citeauthoryear{Knollmann \& Knebe}{2009}]{KK2009}
Knollmann S. R., \& Knebe A., 2009, ApJS, 182, 608

\bibitem[\protect\citeauthoryear{Koester et al.}{2007}]{Koester2007}
Koester, B. P., McKay, T. A., Annis, J., et al. 2007, ApJ, 660, 239

\bibitem[\protect\citeauthoryear{Liao et al.}{2016}]{Liao2016}
Liao S., Gao L., Frenk C. S., Guo Q., \& Wang J. 2016, submitted to MNRAS, arXiv: 1610.07592v1

\bibitem[\protect\citeauthoryear{Mandelbaum et al.}{2010}]{mandelbaum2010}
Mandelbaum R., Seljak U., Baldauf T., \& Smith R. E. 2010, MNRAS, 405, 2078

\bibitem[\protect\citeauthoryear{Mann \& Ebeling}{2012}]{Mann2012}
Mann, A. W., Ebeling, H., 2012, MNRAS, 420, 2120

\bibitem[\protect\citeauthoryear{Li et al.}{2018}]{Li2018}
Li, M.-H., Zhu, W.-S. \& Zhao, D. 2018, arXiv:astro-ph/1805.01165v2.

\bibitem[\protect\citeauthoryear{McConnell et al.}{2011}]{McConnell2011}
McConnell, N. J. et al. 2011, Nature, 480, 215Ð218

\bibitem[\protect\citeauthoryear{Munari et al.}{2013}]{Munari2013}
Munari E., Biviano A., Borgani S., Murante G. \& Fabjan D. 2013,  MNRAS, 430, 2638

\bibitem[\protect\citeauthoryear{Oguri et al.}{2010}]{oguri2010}
Oguri M., Takada M., Okabe N., Smith G. P., 2010, MNRAS, 405, 2215

\bibitem[\protect\citeauthoryear{Rossetti et al.}{2016}]{Rossetti2016}
Rossetti, M., Gastaldello, F., Ferioli, G., Bersanelli, M., De Grandi, S., et al., 2016, MNRAS, 457, 4515

\bibitem[\protect\citeauthoryear{Rozo et al.}{2011}]{rozo2011}
Rozo E., Rykoff E., Koester B., et al. 2011, ApJ, 740, 53

\bibitem[\protect\citeauthoryear{Sanderson et al.}{2009}]{Sanderson2009}
Sanderson, A. J. R., Edge, A. C., \& Smith, G. P., 2009, MNRAS, 398, 1698

 \bibitem[\protect\citeauthoryear{Shan et al.}{2010}]{Shan2010}
Shan H.-Y., Qin B., Fort B., Tao C., Wu X.-P., \& Zhao H.-S. 2010, MNRAS, 406, 1134

\bibitem[\protect\citeauthoryear{Springel}{2005}]{springel2005}
Springel V., 2005, MNRAS, 364, 1105

\bibitem[\protect\citeauthoryear{Viola et al.}{2015}]{viola2015}
Viola M., Cacciato M., Brouwer M., et al. 2015, MNRAS, 452, 3529

\bibitem[\protect\citeauthoryear{Zitrin et al.}{2012}]{zitrin2012}
Zitrin A., Bartelmann M., Umetsu K., Oguri M., \& Broadhurst T., 2012, MNRAS, 426, 2944








\end{thebibliography}
\end{document}